\documentclass[
  aps,
  pre,
  reprint
]{revtex4-2}

\usepackage{amsmath,amssymb,bm}
\usepackage{graphicx}
\usepackage{hyperref}

\begin{document}

\title{Topology-Informed Jet Tagging using Persistent Homology}

\author{Saurav Mittal}
\affiliation{Centre for Development of Advanced Computing (CDAC), Pune, India}

\date{Jan 2026}

\begin{abstract}
We present a topology-informed approach for classifying particle jets using persistent homology, a framework that captures the structural properties of point clouds. Particle jets, produced in proton–proton collisions, consist of cascades of particles originating from a common hard interaction. By representing each jet constituent as a point in a three-dimensional feature space $ X = \{x_i\}_{i=1}^N \subset \mathbb{R}^3, \quad
x_i = (p_{T,i}^{\mathrm{rel}}, \eta_i^{\mathrm{rot}}, \phi_i^{\mathrm{rot}})$  we construct a point cloud description of each jet. Persistent homology is computed using the Vietoris–Rips filtration, yielding persistence diagrams. These diagrams are converted into persistence images for the $H_0$ and $H_1$ homology groups, corresponding to connected components and loop-like structures respectively, which are then used as inputs to a convolutional neural network for quark–gluon jet classification. Using the publicly available HLS4ML jet dataset, we find that  $H_0$ and $H_1$ persistence images individually exhibit comparable discriminative power. These results demonstrate that loop-like topological features, which are not conventionally exploited in jet tagging, encode meaningful information about jet substructure and offer a promising new avenue for particle jet classification.
\end{abstract}

\maketitle

\section{Introduction}

Particle jets are ubiquitous objects in high-energy proton–proton collisions at the Large Hadron Collider. A jet is a collimated spray of particles and serves as a handle to probe the underlying elementary particle produced in the hard scattering process\cite{Qu_2020}. Because of color confinement, quarks and gluons produced in hard interactions hadronize into such collimated sprays, whose internal structure reflects the nature of the initiating particle. For example, gluon jets tend to have a broader energy spread than quark jets \cite{Qu_2020}.

Jet tagging algorithms are designed to identify the nature of the particle that initiated a given cascade, inferring it from the collective features of the particles generated in the cascade\cite{Moreno2020}. This task is crucial for precision Standard Model measurements and searches for new physics.

With the increasing complexity and volume of LHC data, machine-learning and computational approaches have become standard. There is strong motivation to probe the intrinsic shape of jets, a new approach to jet tagging that leverages topological properties of jets to capture their inherent shape\cite{thomas2022topological}.

A concrete realization of this idea is persistent homology. In Jet Topology, Li et al. introduce persistent Betti numbers to characterize topological structure of jets, which measure multiplicity and connectivity of jet branches at a given scale threshold or filtration while their persistence records evolution of each topological feature as this threshold / filtration varies\cite{li2020jettopology}.

In this work, we analyze the publicly available HLS4ML jet dataset \cite{hls4ml}. Each jet is represented as a point cloud, where every point is a jet constituent described by three coordinates: the relative transverse momentum $p_T^{\mathrm{rel}}$, the pseudorapidity $\eta_{\mathrm{rot}}$, and the azimuthal angle $\phi_{\mathrm{rot}}$. Persistent homology is computed on these point clouds using Vietoris-Rips complex filtration, and persistence images for $H_0$, $H_1$ homology group are used as inputs to a convolutional neural network (CNN) for quark–gluon jet classification.

While $H_0$ primarily reflects the multiplicity of jet branches or local density of the point cloud, $H_1$ captures one-dimensional topological features associated with loop-like structures across the filtration as illustrated in Fig.~\ref{fig:1} (a representative one-dimensional feature appears at $K_4$ and disappears after $K_5$). We find that $H_0$ and $H_1$ persistence images individually exhibit comparable discriminative power, and while their combination leads to improved classification performance. This indicates that jet substructure contains complementary information encoded in both connected components and one-dimensional topological features, which can be effectively captured using persistent homology.

While previous studies have shown that topological summaries can be useful for jet characterization, the specific role of one-dimensional features has not been explicitly quantified in a classification setting. This work aims to isolate the contribution of loop-like topological structures to jet tagging.

\section{Mathematical Background and Topological Representation of Jets}

This section introduces the mathematical framework underlying the use of persistent homology
for jet tagging. We describe how particle jets are represented as point clouds, how topological
structures are constructed through a Vietoris--Rips filtration, and how persistent homology
encodes multi-scale geometric information into representations suitable for machine learning.

\subsection{Jets as Point Clouds}

A reconstructed jet consists of a finite collection of particle constituents originating
from a common hard scattering process in a proton--proton collision. Each constituent is
characterized by a set of kinematic variables measured by the detector. To analyze the
internal geometric structure of jets, we represent each jet as a point cloud embedded in a
low-dimensional feature space. 

Concretely, a jet is modeled as a finite set of points

\begin{equation}
X = \{x_i\}_{i=1}^N \subset \mathbb{R}^3, \qquad
x_i = \left(p_{T,i}^{\mathrm{rel}},\, \eta_i^{\mathrm{rot}},\, \phi_i^{\mathrm{rot}}\right),
\end{equation}

where $N$ denotes the number of constituents in the jet. Here $p_T^{\mathrm{rel}}$ is the
transverse momentum of a constituent relative to the jet axis, while $\eta^{\mathrm{rot}}$
and $\phi^{\mathrm{rot}}$ are the pseudorapidity and azimuthal angle defined in a rotated
coordinate frame aligned with the jet axis, as commonly used in jet substructure studies
\cite{Moreno2020}.

This representation treats each jet as an unordered set of points, making it naturally
invariant under permutations of jet constituents\cite{Qu_2020}. Furthermore, persistent homology is
invariant under global translations and rotations of the point cloud, ensuring that the
topological features extracted from the data are insensitive to the choice of coordinate
origin or orientation\cite{book}.

\subsection{Vietoris--Rips Filtration}

To extract topological information from the point cloud $X$, we construct a family of
simplicial complexes using a Vietoris--Rips (VR) filtration, a standard construction in
topological data analysis \cite{Ghrist,book}.

Given a metric $d(\cdot,\cdot)$ on $\mathbb{R}^3$, the Vietoris--Rips complex at scale
$\epsilon > 0$ is defined as
\begin{equation}
\mathrm{VR}_\epsilon(X) =
\left\{
\sigma \subseteq X \;\middle|\;
d(x_i,x_j) \le \epsilon \;\; \forall\, x_i,x_j \in \sigma
\right\}.
\end{equation}

\begin{figure}[]
    \centering
    \includegraphics[width=0.48\textwidth]{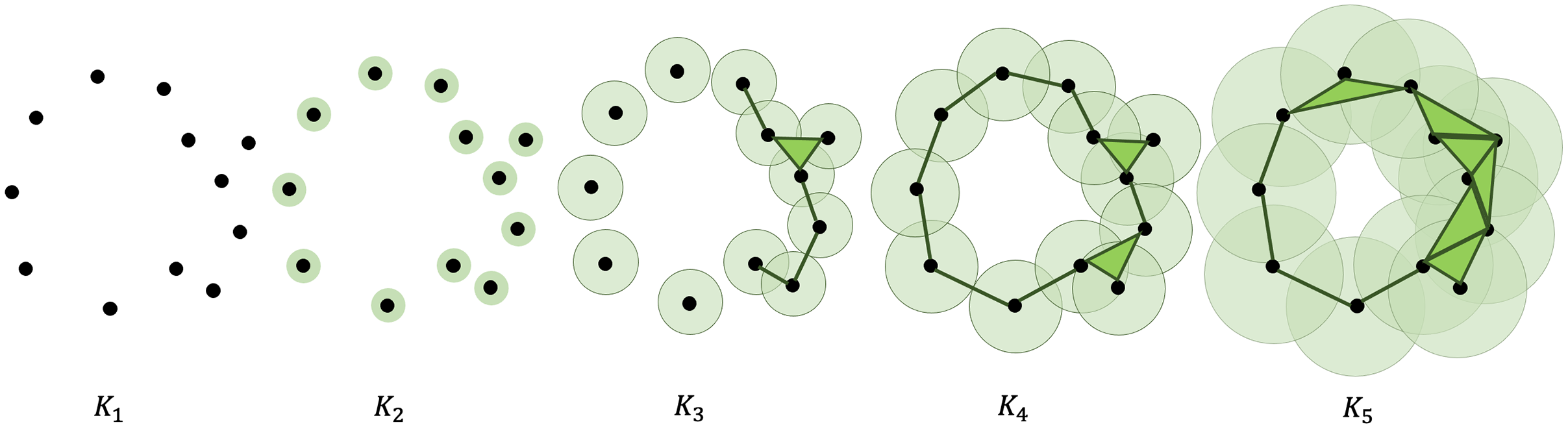}
    \caption{Filtration of a point cloud into a nested sequence of simplicial complexes,
    ${K_1} \subseteq {K_2} \subseteq \dots \subseteq {K_5}$.
    Reproduced from \textit{Dłotko, P. (2022). Understanding Persistent Homology and Its Use in Data Analysis}.
    \textit{PeerJ Computer Science}, 8:e1195, 
    DOI: \href{https://doi.org/10.7717/peerj-cs.1195}{10.7717/peerj-cs.1195}, 
    under a Creative Commons Attribution (CC BY 4.0) license.}
    \label{fig:1}
\end{figure}

In this construction, a $k$-simplex is included whenever all of its $(k+1)$ vertices lie
within pairwise distance $\epsilon$ of one another. As the scale parameter $\epsilon$
increases from zero, isolated points merge into connected components, loops may form, and
higher-dimensional simplices appear, generating a nested sequence of complexes as illustrated 
in Fig.~\ref{fig:1}.
\[
\mathrm{VR}_{\epsilon_1}(X) \subseteq \mathrm{VR}_{\epsilon_2}(X) \subseteq \cdots,
\qquad \epsilon_1 < \epsilon_2 < \cdots .
\]

\subsection{Persistent Homology}

Persistent homology is a central tool within TDA that
identifies topological structures---such as connected components and loops---and quantifies
their persistence across scales, thereby distinguishing robust geometric features from
topological noise \cite{book}.

Given the filtered family of Vietoris--Rips complexes $\{\mathrm{VR}_\epsilon(X)\}$, persistent
homology computes the $k$-th homology group $H_k(\mathrm{VR}_\epsilon(X))$ at each scale
$\epsilon$. The rank of this group,
\begin{equation}
\beta_k(\epsilon) = \mathrm{rank}\, H_k(\mathrm{VR}_\epsilon(X)),
\end{equation}
is known as the $k$-th Betti number. 

Here, $H_k$ denotes the $k$-th homology group, while $\beta_k$ denotes the corresponding Betti number, i.e., the number of topological features (connected components for $k=0$ and loops for $k=1$) contained in $H_k$ at a given filtration scale. The Betti numbers count topological features of different dimensions: $\beta_0$ counts connected components, while $\beta_1$ counts one-dimensional
loop-like structures \cite{Ghrist}. In the context of jet physics, $\beta_0$ reflects 
information about the multiplicity of jet branches or local density of point cloud while 
$\beta_1$ probes loop-like structures in point cloud, which are sensitive to the collective 
spatial organization induced by the underlying hard and subsequent radiative processes.


\begin{figure}[]
    \centering
    \includegraphics[width=0.38\textwidth]{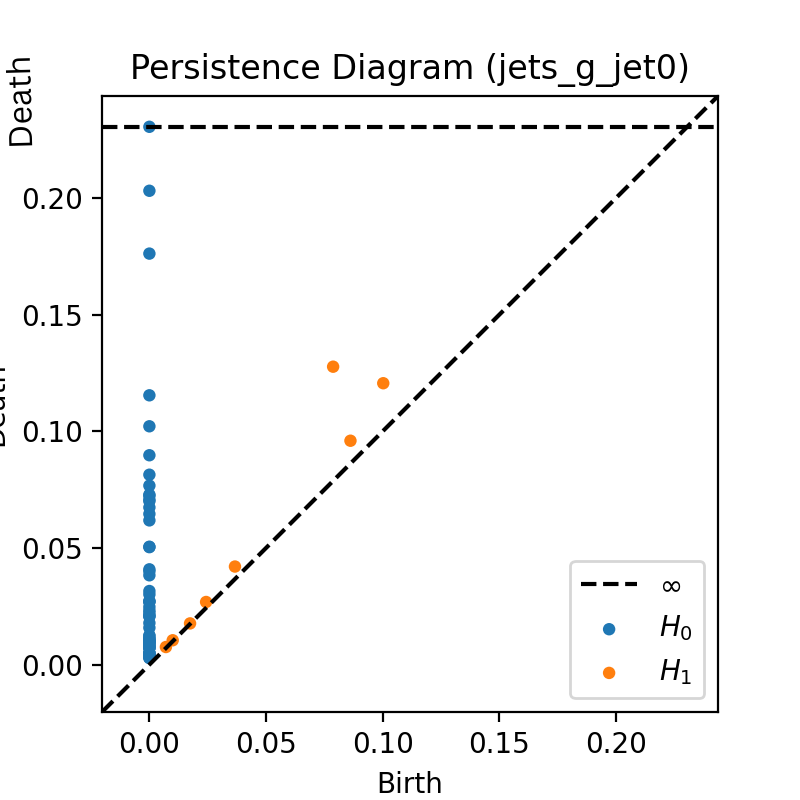}
    \caption{Persistence diagram of a gluon jet on Birth vs Death axis. The perpendicular distance of each topological feature to the diagonal axis represents its respective persistence.}
    \label{fig:2}
\end{figure}

\subsection{Persistence Diagrams and Persistence Images}
\begin{figure}[t]
    \centering

    \begin{minipage}{0.48\columnwidth}
        \centering
        \includegraphics[width=\linewidth]{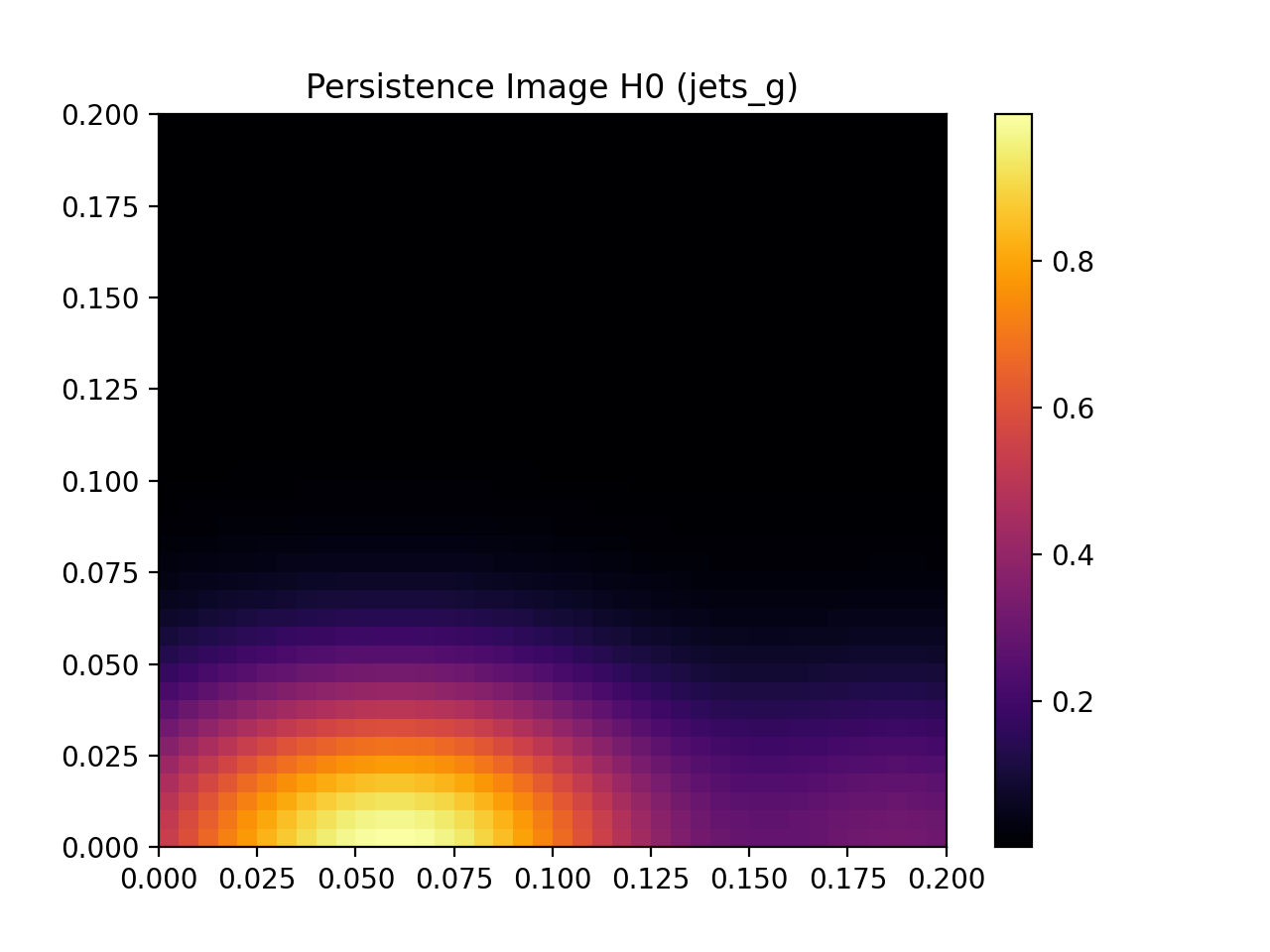}
        \\[-0.5ex]
        (a) Gluon jet $H_0$
    \end{minipage}
    \hfill
    \begin{minipage}{0.48\columnwidth}
        \centering
        \includegraphics[width=\linewidth]{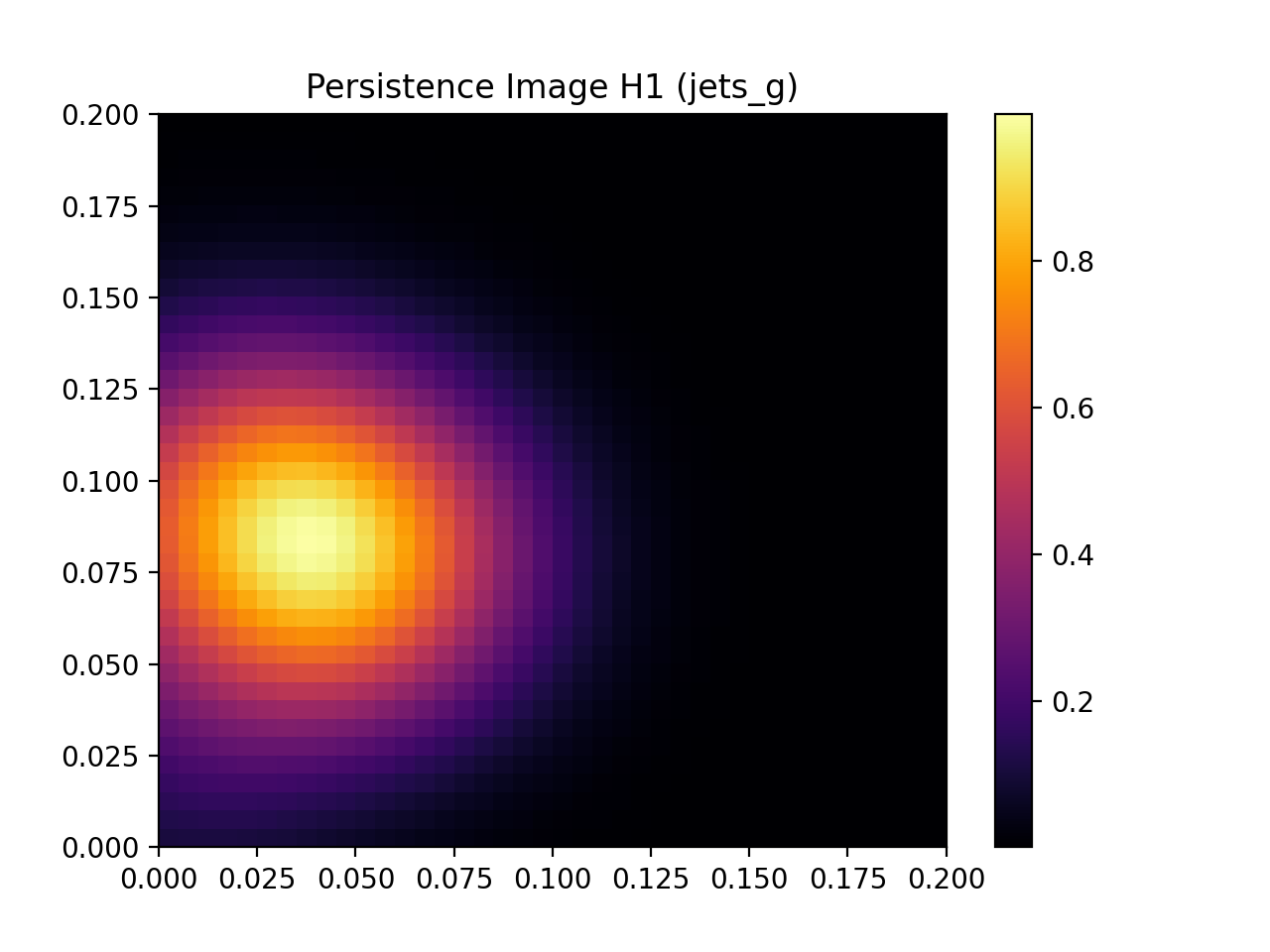}
        \\[-0.5ex]
        (b) Gluon jet $H_1$
    \end{minipage}

    \vspace{0.6em}

    \begin{minipage}{0.45\columnwidth}
        \centering
        \includegraphics[width=\linewidth]{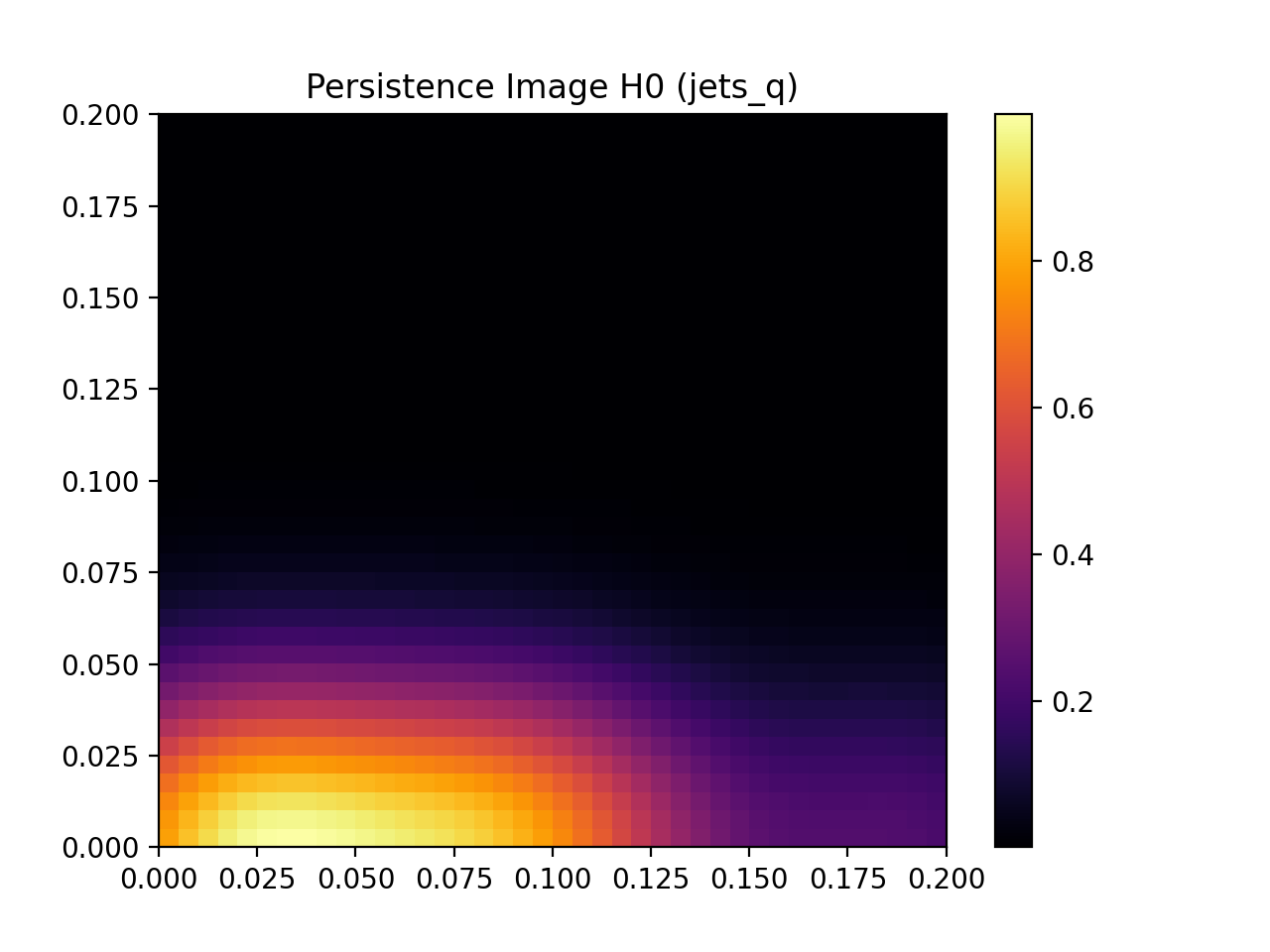}
        \\[-0.5ex]
        (c) Quark jet $H_0$
    \end{minipage}
    \hfill
    \begin{minipage}{0.48\columnwidth}
        \centering
        \includegraphics[width=\linewidth]{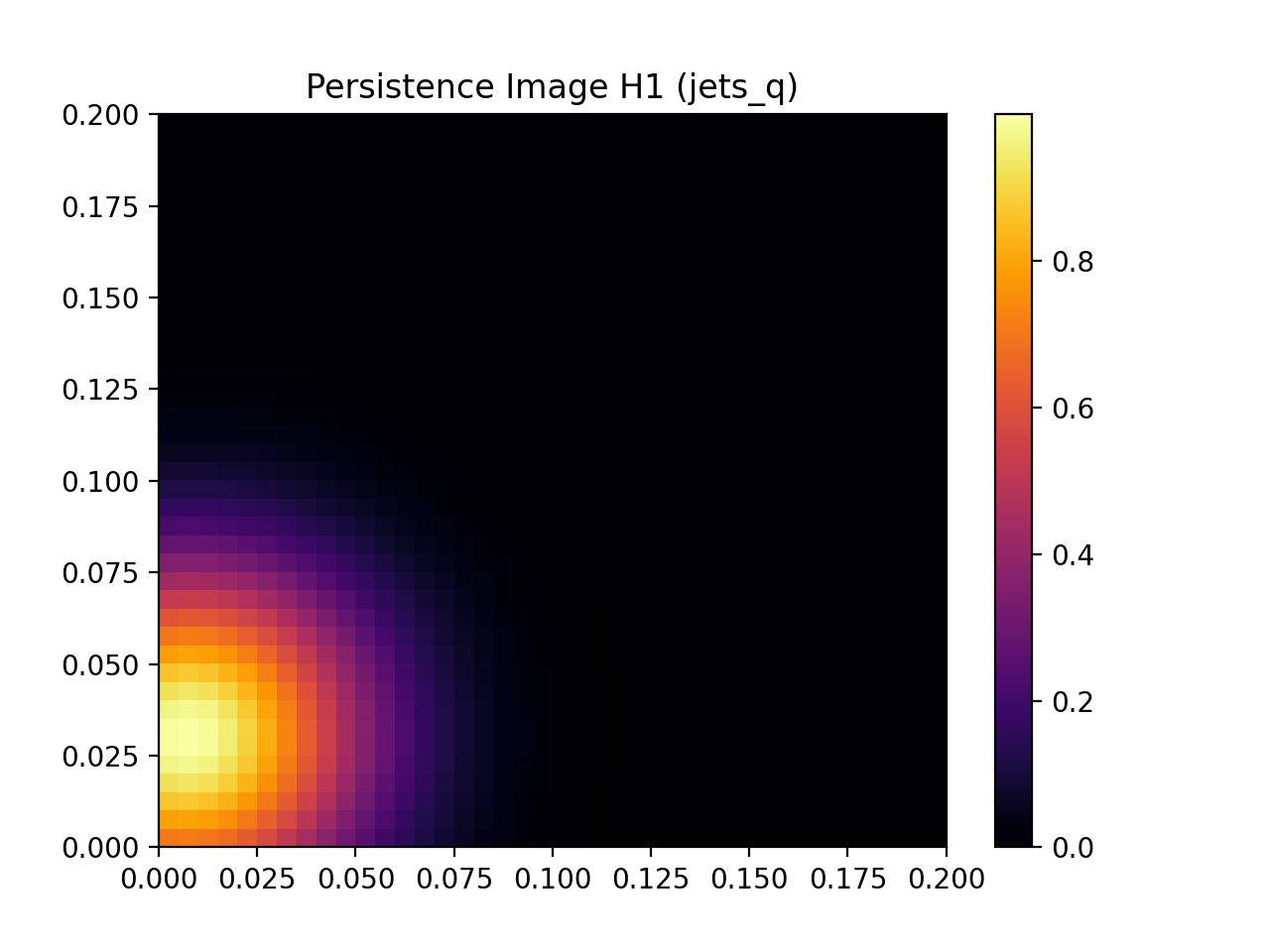}
        \\[-0.5ex]
        (d) Quark jet $H_1$
    \end{minipage}

    \caption{Persistence images constructed from the $H_0$ (connected components) and
    $H_1$ (loops) persistence diagrams for representative gluon- and quark-initiated jets.}
    \label{fig:Persistence_images}
\end{figure}

As the filtration scale $\epsilon$ varies, topological features are created and destroyed, yielding a set of persistence pairs $(b_i, d_i)$, where $b_i$ and $d_i$ denote the birth and death scales of the $i$-th feature, respectively, and its persistence is given by $d_i - b_i$. For a fixed homology dimension, these pairs are summarized in a persistence diagram (see Fig.~\ref{fig:2}), where each point encodes the scale at which a topological feature appears and disappears.

To obtain a fixed-dimensional representation suitable for machine learning, we transform persistence diagrams into persistence images (see Fig.~\ref{fig:Persistence_images}) by placing a Gaussian kernel at each persistence point and discretizing the resulting function on a regular grid~\cite{adams2016persistenceimagesstablevector}. Persistence images preserve the essential multiscale topological information while enabling the use of standard machine learning architectures.

\subsection{Dataset and libraries}
We use the publicly available HLS4ML jet dataset \cite{hls4ml}. Persistent homology computations are performed using the \texttt{ripser} library, and persistence images are generated using the \texttt{persim} package.

\section{Methods and Results}

Jets are represented as point clouds, with each point corresponding to a jet constituent.
For each jet, Vietoris--Rips filtrations were computed using the \texttt{ripser} library to obtain
$\beta_0$ and $\beta_1$ persistence pairs. These diagrams were converted into persistence images using
the \texttt{persim} package with a Gaussian kernel of bandwidth $\sigma = 0.001$ and resolution
$40 \times 40$ for both $H_0$ and $H_1$ homology groups.

These images are used as inputs to a convolutional neural network (CNN) for binary quark--gluon jet
classification, and the performance of the model is compared for $H_0$ and $H_1$ inputs. The CNN
consists of three convolutional blocks with batch normalization and ReLU activations, followed by
fully connected layers and trained using the Adam optimizer. The analysis is performed on a total
of $245{,}109$ jets, with an $80{:}20$ train--validation split.
\begin{figure}[t!]
    \centering
    \includegraphics[width=0.5\textwidth]{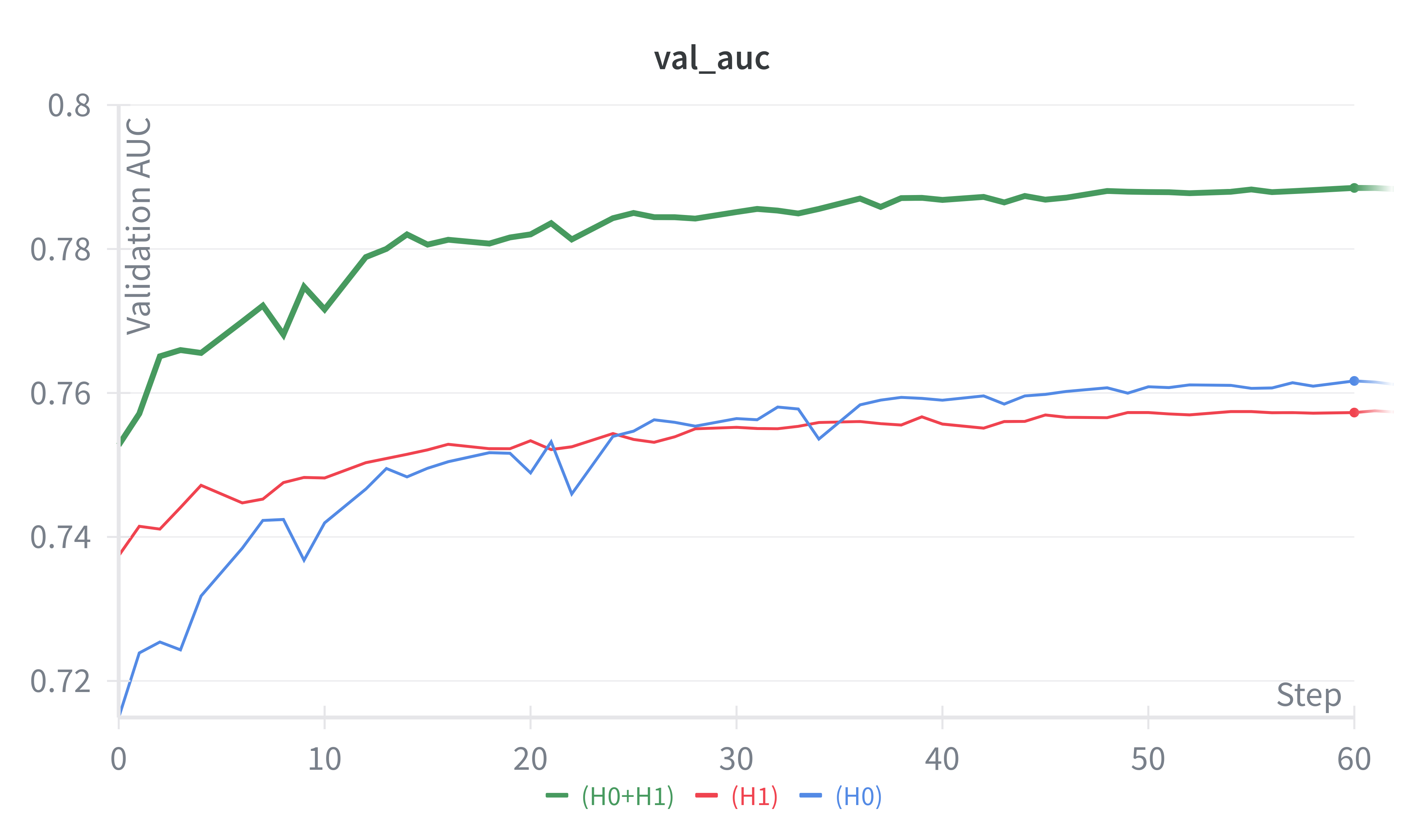}
    \caption{Validation AUC as a function of epoch for the convolutional neural network classifier using $H_0$, $H_1$ and combined persistence image inputs.}
    \label{fig:training}
\end{figure}

We find that persistence images of both homology classes individually exhibit comparable
discriminative power (validation AUC of $0.76$ and $0.75$, respectively) for quark--gluon jet tagging,
while combining the two images gives a slightly better performance, as shown in
Fig.~\ref{fig:training}.

\section{Conclusion}

Persistent homology captures loop-like structures in jets by tracking the emergence and persistence of one-dimensional cycles in the underlying point-cloud representation as the filtration scale varies. The ability of first-order homology to independently discriminate between quark- and gluon-initiated jets indicates that jet substructure encodes physically relevant information in higher-dimensional topology. This observation suggests that jets possess nontrivial geometric organization that is naturally accessed through topological descriptors. As a result, persistent homology provides a principled framework for studying the intrinsic topology of jets, with potential applications to more complex tagging problems and searches for new physics.

\begin{acknowledgments}
The author acknowledges CDAC Pune for the computational resources. This study used the open HLS4ML dataset from CERN Open Data. This project originated during research exposure at IISER Pune.
\end{acknowledgments}

\bibliography{apssamp}

\end{document}